\documentstyle[preprint,aps,graphicx]{revtex}     

\begin{document}      
\title{Collisional dynamics of ultracold OH molecules in an electrostatic field.}      
\author{ Alexandr V. Avdeenkov and John L. Bohn}      
\address{JILA and Department of Physics, University of Colorado, Boulder, CO}      
\date{\today}      
\maketitle      
      
\begin{abstract}      
Ultracold collisions of polar OH molecules are considered in the presence of  
an electrostatic field.  The field exerts a strong influence on both elastic  
and state-changing inelastic collision rate constants, leading to   
clear experimental signatures that should help disentangle the theory  
of cold molecule collisions.  Based on the collision  
rates we discuss the prospects for evaporative cooling of electrostatically   
trapped OH.  We also find that  the scattering properties  
at ultralow temperatures prove to be remarkably independent of the details  
of the short-range interaction, owing to avoided crossings  
in the long-range adiabatic potential curves.  The behavior of the    
scattering rate constants is qualitatively understood in terms of  
a novel set of long-range states of the [OH]$_2$ dimer.  
\end{abstract}

      
\narrowtext      
      
\section{Introduction}      
Polar molecules bring something entirely new to the field of  
ultracold physics.  As compared to the neutral atoms that have  
been studied experimentally in the past, polar molecules possess  
extremely strong, anisotropic interactions.  It has been speculated  
that dipolar interactions will lead to new properties in  
Bose-Einstein condensates \cite{Santos,You,Goral} or degenerate  
Fermi gases \cite{Shlyapnikov}.  It has also been suggested that   
polar molecules in optical lattices may be useful in implementing quantum   
logic elements \cite{DeMille}.  
On the experimental side, cold polar molecules may be produced in  
several ways, including photoassociation of two distinct alkali species  
\cite{Shaffer,Schloder},  
buffer-gas cooling \cite{Doyle,Egorov},   
or Stark slowing \cite{Meijer1,Meijer2,Meijer3}.  
  
Regardless of the method or production, collisions of molecules are  
of paramount importance in describing the properties of the gas.   
Collisions should also be interesting in their own right, as detailed  
probes of intermolecular interactions.  Several features of 
the collisional dynamics of ground-state polar molecules, based on     
a simplified ``toy'' model, were discussed in~\cite{bohnpolar}. This model     
accounted for the interplay between dipole-dipole interactions, an external    
electric field, and states of different parities.  The dipole-dipole    
interaction, which    
scales with intermolecular separation $R$ as $1/R^3$, renders cold    
molecule collisions completely different from cold atom collisions.    
This is because a $1/R^3$ interaction is characterized by energy-independent    
low energy cross sections in {\it all} partial waves, not just in s-waves    
\cite{landau,Shakeshaft,Deb}.  The relatively strong, long-range interactions imply  
that  molecules electrostatically trapped in weak-field-seeking states     
are generally susceptible to state-changing collisions that can      
rapidly deplete the trapped gas.  The rates are in general far larger    
than those of magnetic dipolar transitions in stretched-state  
alkali atoms, owing largely to the    
relative strength of electric, as opposed to magnetic, dipolar    
interactions \cite{bohnpolar}.    
    
In this paper we address ultracold polar molecule collisions in a    
more realistic model, considering in detail the OH radical.  This    
choice is motivated by the attractiveness of this molecule    
for Stark slowing from a supersonic jet \cite{Meijer3,Ye}.    
In particular, it has a $^2\Pi$ ground state with a small $\Lambda$-doublet  
splitting, making it easily manipulated by modest-sized  
electric fields.  
A full treatment of cold collisions is somewhat hindered by    
the fact that the OH-OH potential energy surface (PES) is     
poorly known, although it is known to be very deep and strongly anisotropic    
\cite{kuhn,harding}.  It is not even known, for example, whether    
OH molecules may suffer chemical reactions at ultralow temperatures.    
As a point of reference, it was recently suggested that the reaction   
F+H$_2$ $\rightarrow$  HF + H may proceed at appreciable rates at   
ultralow temperatures,  in spite of having a chemical barrier   
height of 700 K \cite{Bala}.    
    
However, long-range dipole-dipole    
forces strongly dominate the scattering of OH molecules in their    
weak-field-seeking states.  In this paper we will show that this arises    
from strong avoided crossings in the long-range adiabatic potential    
curves, which prevent the molecules from approaching close enough to  
one another for exchange potentials to become important.    
In this regard cooling and electrostatic trapping of     
OH molecules can provide a wealth of information on the long-range    
OH-OH interaction.  Thus it appears possible to understand   
a class of ultracold    
OH-OH collisions without detailed knowledge of the short range PES.    
This strategy would be an important stepping stone toward understanding    
the full problem of ultracold OH collisions.  Strong-field seekers,    
by contrast, will in addition experience the short-range     
interaction.  The complete problem of exploring collisions of ultracold    
polar molecules might therefore most efficiently proceed by a two-step analysis,    
thus simplifying this very complicated problem.    
    
Accordingly we focus in this paper on the first step, namely, collisions of    
weak-field-seeking states.  After some discussion of the relevant    
properties of OH molecules in Sec. II and their interactions in Sec. III,     
we move on in Sec. IV to illustrate some prominent energy- and    
field-dependent features in elastic and inelastic cross sections.    
Mapping these features in experiments should help in unraveling     
the long-range part of this puzzle.  We also present a simplified    
model of the long-range interaction, to help illustrate the    
basic physics behind the behavior of the cross sections.  It will  
turn out that a new class of long-range bound states of the [OH]$_2$  
dimer play a significant role in ultracold collisions of this  
molecule.  
    
\section{OH molecule}      
    
The OH molecule has a fairly complicated internal structure,     
incorporating rotation, parity, electronic spin, and nuclear spin    
degrees of freedom, which are  further confounded in    
the presence of an electric field.  We therefore begin by     
describing the structure of this molecule and the simplifications    
we impose to render our model tractable.    
    
    
Molecules cooled to sub-Kelvin temperatures by Stark slowing will be    
assumed to be in their electronic $^{2}\Pi$ ground state,      
and $\upsilon =0 $ vibrational ground state.  In this state OH is an   
almost pure Hund's case (a) molecule, and has a dipole moment of   
1.68 D \cite{kuhn}.    
Spin-orbit coupling involving the lone electronic spin     
splits the ground state into $^{2}\Pi_{3/2}$ and $^{2}\Pi_{1/2}$ components,  
of which $^{2}\Pi_{3/2}$ is lower in energy and is therefore the    
state of greatest interest in ultracold collisions.    
In our model we take into account just the lowest rotational      
level of the corresponding ground state, $J=3/2$.     
The energy of the first rotationally excited state with $|J=5/2,\Omega=3/2>$     
is about 84K higher in energy \cite{coxon} and we will neglect this and  
higher-lying states     
in our scattering calculations.   Such states will, however, contribute  
rotational Feshbach resonances in realistic collisions.  
     
The isotopomer $^{16}$O$^1$H that we consider here has a nuclear spin      
of $I=1/2$, which with a half- integer rotational quantum      
number  defines the OH molecule as a boson. Thus  we should take into      
account hyperfine structure  to ensure the proper Bose symmetry.    
We will see below that the inclusion of hyperfine structure is  
also important in determining details of collision properties.  
The calculations in an electric field also require  knowing the Stark splitting     
for OH molecules.   Thus the Hamiltonian for the OH molecule in a field is      
\begin{equation}      
\label{hamone}      
H^{OH}=H_{rot}+H_{fs}+H_{hfs}+H_{field}    
\end{equation}      
    
Wave functions for the spatial degrees of freedom of the molecule    
are constructed in the usual way.  Namely, in the zero-electric-field    
limit, eigenstates of parity $\varepsilon$ ($= \pm$) are given by the Hund's case (a)    
representation:    
\begin{equation}      
\label{wfs}      
|J M_{J} \Omega  \varepsilon>=\frac{1}{\sqrt{2}} \bigl(      
|J M_{J} \Omega > |\Lambda \Sigma> + \varepsilon |J M_{J} -\Omega > |-\Lambda -\Sigma>      
\bigr),      
\end {equation}      
where the rotational part is given by a Wigner function:      
\begin{equation}      
|J M_{J} \Omega > =(\frac{2J+1}{8\pi^{2}})^{1/2} D^{J}_{M_{J} \Omega}(\theta,\phi,\kappa)  ,    
\end{equation}      
and $\Omega=|\Sigma+ \Lambda|$ is the projection of the total  electronic    
angular momentum on the molecular axis.    
The total spin of the molecule, ${\bf F}= {\bf J} +{\bf I}$, with laboratory    
projection $M_{F}$, is then constructed by    
\begin{equation}      
\label{onestate}      
|(JI)FM_{F}\Omega \varepsilon>=|\Lambda> |S \Sigma>      
\sum_{M_{J},M_{I}} |J M_{J} \Omega  \varepsilon> |IM_{I}><FM_{F}|JM_{J}I_{M_{I}}>,      
\end{equation}      
    
The matrix elements for the Hamiltonian (\ref{hamone}) in this basis    
can be found elsewhere~\cite{mizushima}. In compact form these matrix      
elements are    
\begin{eqnarray}      
\label{matrix1}      
<(JI)F \Omega M_{F} \varepsilon| H^{OH}|(J'I')F' \Omega' M_{F'} \varepsilon'>=\Bigl(      
\delta_{\Omega,3/2} \delta_{\Omega',3/2} E_{3/2,3/2}+      
\\      
\nonumber      
\delta_{\Omega,1/2} \delta_{\Omega',1/2} E_{1/2,1/2}+      
(\delta_{\Omega,3/2} \delta_{\Omega',1/2}+      
\delta_{\Omega,1/2} \delta_{\Omega',3/2} E_{3/2,1/2}      
\Bigl) \times      
\\      
\nonumber      
\delta_{J,J'} \delta_{F,F'} \delta_{\varepsilon,\varepsilon'}-      
\mu {\cal E} \frac{1}{2}(1+(-1)^{J+J'}\varepsilon \varepsilon')(-1)^{F+F'+M_{F}+I-\Omega+1}      
\times      
\\      
\nonumber      
([J][J'][F][F'])^{1/2}      
\left( \begin{array}{ccc}      
                  J & 1 & J' \\      
                  -\Omega & 0 & \Omega'      
                  \end{array} \right)      
\left( \begin{array}{ccc}      
                  F' & 1 & F \\      
                  -M_{F'} & 0 & M_{F}      
                  \end{array} \right)      
\left\{ \begin{array}{ccc}      
                  1 & J & J' \\      
                  I & F & F'      
                  \end{array} \right\}.    
\end{eqnarray}      
In this expression $\mu$ is the molecular dipole moment,     
${\cal E}$ is the strength of the electric field, and      
$E_{\Omega,\Omega^{\prime}}$ are matrix elements for      
the fine structure   
$H_{rot} +H_{fs}$ which  can be found in~\cite{mizushima,miller}.    
These values depend on the rotational constant, spin-orbit coupling constant,      
hyperfine coupling constant, and $\Lambda$-doublet parameters of OH.    
All of these constants can be found in~\cite{coxon}.     
  
Equation (\ref{matrix1}) shows that, strictly speaking, the only  
good quantum number for the OH molecule is the projection of its  
angular momentum on the laboratory axis, $M_F$.  However, for our  
present purposes it suffices to treat the quantum numbers as  
``almost good.''  For example, in view of the fact that OH is nearly  
a purely Hund's case (a) molecule, the coupling between $\Omega = 1/2$  
and $\Omega = 3/2$ states is fairly weak. We account for this interaction  
perturbatively, by replacing the values $E_{3/2}$ and $E_{1/2}$  
by the eigenvalues of the $2 \times 2$ matrix  
\begin{equation}  
\left( \begin{array}{cc}  
  E_{3/2} & E_{3/2,1/2} \\  
  E_{3/2,1/2} & E_{1/2} \\  
\end{array} \right),  
\end{equation}   
keeping all other quantum numbers constant.  
  
Likewise, different values of the molecular spin $J$ are mixed in a field,  
but this mixing is small in laboratory strength fields.  The total spin  
$F$ and the parity are far more strongly mixed.  Accordingly,  
in practice we transform the molecular state to a field-dressed basis  
for performing scattering calculations:  
\begin{eqnarray}     
\label{ebasis}     
|(\tilde{J}I)\tilde{F}M_{F} \Omega \tilde{\varepsilon};{\cal E} > \equiv  
\sum_{JF\varepsilon} \alpha(JF\varepsilon) |(JI)FM_{F}\Omega \varepsilon>,     
\end{eqnarray}     
where $\alpha(JF\varepsilon)$ stands for eigenfunctions of the Hamiltonian~(\ref{hamone})    
determined numerically at each value of the field.  We will continue    
to refer to molecular states by the quantum numbers $J$, $F$, and $\varepsilon$,   
with the understanding that they are only  approximately good in a field,  
and that (\ref{ebasis}) is the appropriate molecular state.  
    
Figure 1 shows the Stark energies computed using all the ingredients  
described above. In zero field the energy levels are primarily determined     
by the $\lambda$-doublet splitting between opposite parity states,    
whose value is $\Delta=0.056cm^{-1}$.  The alternative parity states,  
with $\varepsilon=-1$ (f states) and $\varepsilon=+1$ (e states) are  
shown in Fig. 1(a) and Fig. 1(b), respectively.  These states are further split    
into hyperfine components with total spin $F=1$ and $F=2$.  The Stark shift    
is quadratic for fields below the critical field     
${\cal E}_{0}\equiv{\Delta}/{2\mu}$ ($\approx 1000 V/cm$ for OH).    
For fields larger than ${\cal E}_{0}$ states with different parity are entirely      
mixed and the Stark effect transforms from quadratic to linear.     
In this case the molecular states are roughly equal linear combinations  
of the zero-field $\varepsilon=-$ and $\varepsilon=+$ states (compare  
Eqn. \ref{wfs}) \cite{schreel}:  
\begin{eqnarray}      
|J M_{J} \Omega  \varepsilon=\pm 1> = \{ \begin{array}{cc}      
                                        | J M_{J} \mp \Omega > &  M_{J} < 0 \\      
                                        | J M_{J} \pm \Omega > &  M_{J} > 0      
                                           \end{array}      
\end{eqnarray}      
     
\section{OH-OH Interaction}      
We will consider diatom-diatom scattering as two interacting      
rigid rotors in their ground rotational states.     
The complete Hamiltonian for the collision process can then be written     
\begin{equation}      
H=T_{1}+T_{2}+H^{OH}_{1}+H^{OH}_{2}+V_{s}+V_{\mu \mu}+V_{qq}+V_{disp},   
\end{equation}      
where $T_{i}$ and $H^{OH}_{i}$ are the translational kinetic energy     
and internal motion of molecule $i$,  including the electric field    
as in Eqn. (\ref{hamone}). $V_{s}$ is the short- range exchange      
interaction, $V_{\mu \mu}+V_{qq}+V_{disp}$ are the dipole- dipole,      
quadrupole- quadrupole and dispersion long- range interactions      
respectively.  Explicit expression for the dipole-dipole ($\propto 1/R^3$)  
and quadrupole-quadrupole ($\propto 1/R^5$) interactions are given  
in Ref. \cite{avoird}.  Matrix elements for the the dipole-quadrupole   
interaction vanish for rigid rotor molecules in identical states
\cite{kuhn}, hence will not  be considered here.  
    
The anisotropic potential between two interacting rigid-rotor molecules    
is conveniently recast into a standard set of angular functions~\cite{avoird}:      
\begin{eqnarray}      
\label{potential}      
V_{s}+V_{\mu \mu}+V_{qq}+V_{disp} \equiv     
V(\omega_{A},\omega_{B},\omega,R)=\sum_{\Lambda} V_{\Lambda} A_{\Lambda}(\omega_{A},\omega_{B},\omega),      
\end{eqnarray}      
where $\Lambda \equiv (L_{A},K_{A},L_{B},K_{B},L)$ and the angular      
functions are defined as:      
\begin{eqnarray}      
A_{\Lambda}(\omega_{A},\omega_{B},\omega)= \sum_{M_{A},M_{B},M}      
\left( \begin{array}{ccc}      
                  L_{A} & L_{B} & L \\      
                  M_{A} & M_{B} & M      
                  \end{array} \right)      
D^{L_{A}}_{M_{A},K_{A}}(\omega_{A})     
D^{L_{B}}_{M_{B},K_{B}}(\omega_{B})     
C^{L}_{M}(\omega),      
\end{eqnarray}      
where $\omega_{A,B}=(\theta_{A,B}, \phi_{A,B})$ are the polar angles   
of molecules A and B with respect to the lab-fixed quantization axis, and     
${\bf R}=(R,\omega)$ is the  
vector between the center of mass of the molecules in the      
laboratory fixed coordinate frame.  The indices $K_A$ and $K_B$    
denote the dependence of the interaction on the orientation of    
the molecules about their own axes; in what follows we will ignore    
this dependence, setting $K_A=K_B=0$.  For the long-range part    
of the interaction this approximates the quadrupole moment of OH    
as cylindrically symmetric.      
    
The exchange potential $V_s$ is very complicated, consisting of    
four singlet and four triplet surfaces \cite{harding}, and is moreover  
poorly characterized.  The most complete treatment of this surface to    
date computes the lowest-energy potential for each value of    
internuclear separation $R$ \cite{kuhn}.  This potential finds    
an extremely deep minimum at $R=2.7$ a.u. corresponding to chemically    
bound hydrogen peroxide, and a second, shallower minimum at $R=6$ a.u.   
due to hydrogen bonding forces.    However, in cold    
collisions, the scattering cross sections are so sensitive to    
details of the short-range interaction that knowing the complete interaction    
probably would not help anyway.  More importantly, as we will see below,   
collisions of the weak-field-seeking states are strongly dominated    
by the long-range dipole-dipole interaction.  Therefore, we will use    
at small $R$ simply the hydrogen-bonding part of the potential    
surface (see Fig. 13 of Ref. \cite{kuhn}), and we will treat this    
part of the interaction as if it were isotropic.  
Finally, we will assert that the spin states of the OH molecules    
are in their stretched states, so that ordinary spin-exchange processes  
will  not play a role in these collisions.    
    
We express the Hamiltonian in a basis of projection of total      
angular momentum,      
\begin{eqnarray}      
\label{wave}      
{\cal M}= M_{F_{1}}+M_{F_{2}}+M_{l},      
\\      
M_{F_{i}}=M_{J_{i}}+M_{I_{i}},      
\end{eqnarray}      
where $M_{F_{i}}, M_{J_{i}}, M_{I_{i}}$ are the projections of the      
full molecule spin, rotational motion and nuclear spin on the      
laboratory axis respectively for each molecule. $M_{l}$ is the      
projection of the partial wave quantum number on the laboratory      
axis. In this basis the wave function for two molecules  is     
described  as:     
\begin{eqnarray}      
\label{basis}      
\Psi^{\cal M}=\sum_{1,2,l,M_{l}} \left\{|1> \otimes |2> \otimes |l M_{l}> \right\}^{\cal M} \times \psi^{{\cal M},1,2}(R),      
\end{eqnarray}      
where $\left\{...\right\}^{\cal M}$ is angular momentum  part of      
this wave function and $|i>$ is the wave function for each      
molecule as described by Eq.(\ref{onestate}).      
      
Because the target and the projectile are identical bosons, we must take   
into account the symmetry of the wave function under exchange. The    
properly symmetrized wave function is then    
\begin{eqnarray}      
\label{swave}      
\left\{|1> \otimes |2> \otimes |l M_{l}> \right\}^{s}=\frac{\left\{|1> \otimes |2> \otimes |l M_{l}> \right\}      
+(-1)^{l} \left\{|2> \otimes |1> \otimes |l M_{l}> \right\}}{\sqrt{2(1+\delta_{12})}}      
\end{eqnarray}      
      
Using the expansion of the intermolecular      
potential~(\ref{potential}), the wave function~(\ref{wave}), and      
taking into account the Wigner-Eckart theorem, we can present the      
reduced angular matrix element as      
\begin{eqnarray}      
\label{matrix2}      
<12lM_{l}||| A_{\Lambda}|||1'2'l'M_{l'}>=      
\\      
\nonumber      
(-1)^{L_{A}+L_{B}+J_{1}+J_{1}'+J_{2}+J_{2}'+M_{F_{1}}' + M_{F_{2}}'-\Omega_{1}'-\Omega_{2}'+M_{l}-1}      
\frac{(1+\varepsilon_{1}\varepsilon_{1}'(-1)^{L_{A}})}{2}      
\frac{(1+\varepsilon_{2}\varepsilon_{2}'(-1)^{L_{B}})}{2}      
\times      
\\      
\nonumber      
([l][l'][J_{1}][J_{1}'][J_{2}][J_{2}'][F_{1}][F_{1}'][F_{2}][F_{2}'])^{1/2}      
\left( \begin{array}{ccc}      
                  L_{A} & L_{B} & L \\      
                  M_{F_{1}}-M_{F_{1}'} &  M_{F_{2}}-M_{F_{2}'} &  M_{l}-M_{l'}      
                  \end{array} \right)      
\times      
\\      
\nonumber      
\left( \begin{array}{ccc}      
                  J_{1}' & L_{A} & J_{1} \\      
                  \Omega_{1}' &  0 &  -\Omega_{1}      
                  \end{array} \right)      
\left( \begin{array}{ccc}      
                  J_{2}' & L_{B} & J_{2} \\      
                  \Omega_{2}' &  0 &  -\Omega_{2}      
                  \end{array} \right)      
\left( \begin{array}{ccc}      
                  L_{A} & F_{1} & F_{1}' \\      
                  M_{F_{1}}-M_{F_{1}'} & -M_{F_{1}}&  M_{F_{1}'}      
                  \end{array} \right)      
\times      
\\      
\nonumber      
\left( \begin{array}{ccc}      
                  L_{B} & F_{2} & F_{2}' \\      
                  M_{F_{2}}-M_{F_{2}'} & -M_{F_{2}}&  M_{F_{2}'}      
                  \end{array} \right)      
\left( \begin{array}{ccc}      
                  l' & L & l \\      
                  M_{l'} & M_{l}-M_{l'} &  -M_{l}      
                  \end{array} \right)      
\left( \begin{array}{ccc}      
                  l' & L & l \\      
                  0 & 0 &  0      
                  \end{array} \right)      
\times      
\\      
\nonumber      
 \left\{ \begin{array}{ccc}      
                  L_{A} & F_{1} & F_{1} \\      
                  I & J_{1'} &  J_{1}      
                  \end{array} \right\}      
 \left\{ \begin{array}{ccc}      
                  L_{B} & F_{2} & F_{2} \\      
                  I & J_{2'} &  J_{2}      
                  \end{array} \right\},      
\end{eqnarray}      
where $I=1/2$ is each molecule's nuclear spin.     
      
The reduced matrix elements of the angular functions   
$A_{\Lambda}$ between symmetrized      
basis states~(\ref{swave}) are      
\begin{eqnarray}      
<12lM_{l}||| A_{\Lambda}^{s}|||1'2'l'M_{l'}>=      
\frac{<12lM_{l}||| A_{\Lambda}^{s}|||1'2'l'M_{l'}>+      
(-1)^{l}<21lM_{l}||| A_{\Lambda}^{s}|||1'2'l'M_{l'}>}      
{\sqrt{(1+\delta_{1,2})(1+\delta_{1',2'})}}      
\\      
\nonumber      
\times      
\frac{1+(-1)^{l+l'}}{2}      
\end{eqnarray}      
In practice, before each scattering calculation we transform the    
Hamiltonian matrix from this basis into the field-dressed basis    
defined by (\ref{ebasis}).    
We solve the coupled- channel equations using a logarithmic  
derivative propagator method \cite{Johnson} to      
determine scattering matrices. Using these matrices we      
calculate total state-to-state cross sections and rate constants.

\section{Results}      
This paper considers the scattering problem for OH molecules      
in an electrostatic field for cold and ultracold temperatures.     
We are interested in particular in the highest energy weak-field-seeking   
state of the ground rotational state,    
$|(J,I)FM_{F},\Omega  \varepsilon>=|(3/2,1/2)22,3/2,->$.  This state is   
indicated by the heavy solid line in Figure 1.  
Since the quantum numbers $J$, $I$, and $\Omega$ are the same for    
all the scattering processes we will consider, we will refer to this    
state by the shorthand notation $|FM_F, \varepsilon \rangle$ $= |22,-> $.    
    
 The main novel feature of OH- OH scattering, as    
compared to atoms or nonpolar molecules, is the presence of    
the long-range dipole-dipole interaction and its dependence on    
the electrostatic  field.   Because these interactions strongly    
mix different partial waves, it is essential that we include    
more than one value of $l$.  However, in the interest of emphasizing  
the basic underlying physics, we have included only    
the s- and d- partial waves.  Sample calculations show that
higher partial waves change the results only slightly at the
energies considered.  In this case, given the initial    
state with $M_{F1} = M_{F2} = 2$, the only allowed values of    
the total projection are ${\cal M} = 2,3,4,5,6$.  Among these    
channels only the one with ${\cal M}=4$ contains a contribution    
from s-wave scattering, and so will deserve special attention     
in what follows.  In this case the total number of scattering channels  
for all allowed values of ${\cal M}$, is 208.    
     
\subsection{Prospects for evaporative cooling}    
  
One of the goals of the present work is to revisit the conclusions    
of Ref. \cite{bohnpolar} concerning the effectiveness of evaporative    
cooling for electrostatically trapped molecules.  To this end    
Figure 2 plots the elastic and state-changing inelastic rate constants versus    
field strength for two different collision energies, 100 $\mu K$ and 1$\mu K$.      
Here ``elastic'' refers to collisions that do not change the internal    
state of either molecule, while ``inel'' denotes those collisions    
in which one or both molecules are converted into any other   
states.  These transitions are typically exothermic, leading to trap  
heating.  Not all these collisions produce untrapped states, however.  
We find that the main contributions to the $K_{inel}$ are given    
by processes in which  quantum numbers $F$ and/or $M_{F}$ are changed by one.   
In particular, the process $|22,- \rangle + |22,- \rangle$
$\rightarrow$ $|22,->+|21,->$  generally makes the  
largest contribution to $K_{inel}$, especially at high electric field.  
  
At low field the rates are nearly independent of field, but begin to evolve    
when the field approximately exceeds the critical field ${\cal E}_0=\Delta/2\mu$    
where the Stark effect changes from quadratic to linear.  Above    
this field the rate constants exhibit oscillations as a function    
of field.  These oscillations provide an experimentally variable  
signature of resonant collisions, meaning that mapping this field  
dependence  should help in untangling details of the long-range  
OH-OH interaction.  This is similar to the ability of magnetic-field  
Feshbach resonances in the alkali atoms to yield detailed scattering  
parameters \cite{Roberts,Loftus}.  
    
Following the example of ultracold atoms, we expect that evaporative    
cooling can proceed when the ratio of elastic to  inelastic collisions,    
$K_{el}/K_{inel} \gg 1$.  Figure 2    
shows that this is hardly ever the case for large field values    
${\cal E} > {\cal E}_0$, except, perhaps, at very special field    
values where $K_{inel}$ is at a minimum of its oscillation.     
Since the losses are dominated by exothermic processes,    
the ratio $K_{el}/K_{inel}$ in the threshold scattering limit    
scales as the ratio $k_i / k_f$ of the incident and final channel    
wave numbers, as can be seen from the Born approximation.  
Thus at high electric fields, where the Stark splitting  
is large (hence $k_f$ is large), the ratio may become more    
favorable.  In our calculations this apparently happens for    
fields above $10^{4}$ V/cm.    
    
For fields below ${\cal E}_0 \approx 1000$ V/cm, a favorable ratio    
of $K_{el}/K_{inel}$ is only somewhat more likely.  For fields this low,   
however, the maximum depth of an electrostatic trap is $\approx 8$mK,    
as given by the magnitude of the Stark shift (Fig. 1).  The temperature of    
the trapped gas must therefore be well below this temperature.  In the  
example of a 100 $\mu$K gas, (Fig. 2a), the ratio $K_{el}/K_{inel}$  
may indeed be favorable.    
However, if the gas is cooled further, say to 1 $\mu$K (Fig. 2b),    
this ratio becomes less favorable again.  This is because of the Wigner    
threshold laws: the exothermic rate $K_{inel}$ is energy-independent    
at low energy, while the elastic scattering rate plummets to zero    
as the square root of collision energy.  Thus, in general, evaporative  
cooling seems to be viable only over an extremely limited range of temperature    
and field  
range for the OH molecule, if at all.  We therefore reiterate the message of    
Ref. \cite{bohnpolar}, and recommend that cold OH molecules    
be trapped by a far-off-resonance optical dipole trap, in their lowest  
energy $|F|M_F| , \varepsilon \rangle = |11,+ \rangle$ states.  
  
At this point we emphasize an essential difference between evaporative  
cooling of electrostatically trapped polar molecules and of magnetically  
trapped paramagnetic molecules.  For polar molecules the transition   
from weak- to strong-field seeking states is {\it always} exothermic,  
even in zero applied field.  This is because the lower  
member of a $\Lambda$-doublet is always strong-field-seeking (e.g. Fig. 1).  
For paramagnetic molecules, by contrast, weak- and strong-field  
seeking states can be nearly degenerate at low magnetic field values  
(e.g., $^{17}$O$_2$ as discussed in Refs. \cite{Avdeenkov,Volpi}.  
In the present case, OH is also paramagnetic and hence could in principle  
be magnetically trapped.  For example, the low-energy states with  
$|FM_F \varepsilon \rangle$ $= |11, + \rangle$  
might be suitable candidates.  The influence of electric dipole  
interactions on evaporative cooling of magnetically trapped OH has  
yet to be explored.

\subsection{Analysis of the long-range interaction}      
    
The general behavior of the rate constants in Fig. 2 can be    
explained qualitatively by simplifying our model even further,    
to a case that contains only the essential ingredients:     
the dipole-dipole interaction, the $\Lambda$-doublet, and an electric    
field \cite{bohnpolar}.  Roughly speaking, the electric field has two effects    
on the molecules: 1) it mixes molecular states of opposite    
parity, thus creating induced dipole moments; and 2) the    
resulting dipole-dipole interaction strongly couples     
scattering channels with different partial waves, leading    
to long-range couplings between two molecules.   
   
As a starting point in this analysis, 
Figure 3 breaks down the elastic and inelastic rates into their    
contributions from different values of the total projection     
of angular momentum ${\cal M}$.  This is done for the rates    
calculated at an energy $E = 100$ $\mu$K, from Figure 2(a).      
In both elastic and inelastic scattering, the rates are dominated    
by the contribution from ${\cal M}=4$, which, it will be recalled,    
is the only value of ${\cal M}$ that incorporates s partial waves  
in the present model.  
We will accordingly consider only this case in what follows.    
    
The model used to obtain the results in Figs 2 and 3 consists    
of 32 channels for the block of the Hamiltonian matrix with    
${\cal M}=4$.  To simplify the analysis of this block even further,    
we focus on the sub-Hamiltonian with fixed quantum numbers    
$F = M_F = 2$ for each molecule.  This reduces the effective    
Hamiltonian to six channels:  there are three non-degenerate    
thresholds $E_{\varepsilon_1 \varepsilon_2}$ corresponding     
to different possible values of the field-dressed parity quantum number  
$\varepsilon_i = \pm$.    
For each of these three thresholds there are two channels, corresponding   
at large $R$ to s- and d- partial waves.    
    
The simplified six-channel Hamiltonian matrix then consists of     
$3 \times 3$ blocks $\hat{V}^{ll^{\prime}}$ parameterized     
by partial wave quantum numbers $l$, $l^{\prime}$:    
\begin{eqnarray}    
\label{model}    
\hat{H} =   
\left( \begin{array}{cc}    
                  \hat{V}_{diag}^{00} &   \hat{V}_{\cal E}^{02} \\    
                  \hat{V}_{\cal E}^{20} & \hat{V}_{diag}^{22}+\hat{V}_{\cal E}^{22}    
                  \end{array} \right)    
\end{eqnarray}    
Here the diagonal components $\hat{V}_{diag}^{ll^{\prime}}$ include    
the parity thresholds and the centrifugal interactions,    
\begin{eqnarray}    
\label{modeldiag}    
\hat{V}_{diag}^{ll}=    
\left( \begin{array}{ccc}    
                  E_{--}+{\hbar^2 l(l+1) \over 2m R^2} & 0 & 0  \\    
                  0 & E_{-+}+{\hbar^2 l(l+1) \over 2m R^2} & 0 \\    
                  0 &  0 & E_{++}+{\hbar^2 l(l+1) \over 2m R^2}    
                  \end{array} \right)    
\end{eqnarray}    
where the electric-field-dependent thresholds are given by    
$E_{\varepsilon_{1}\varepsilon_{2}}    
=E_{+}+E_{-}-({\varepsilon_{1}+\varepsilon_{2}}) \Delta    
\sqrt{1+k^{2}}/2$, in terms of the dimensionless parameter   
\begin{equation}  
\label{kdef}  
k \equiv {2 \mu{\cal E} \over \Delta}  
\end{equation}  
that relates the electric    
field strength ${\cal E}$ to the zero-field lambda-doublet splitting   
$\Delta = E_- - E_+$.    
The simplified field-dependent dipole-dipole interaction term     
$\hat{V}_{{\cal E}}^{ll^{\prime}}$  is readily parameterized in the   
field-dressed  basis as   
\begin{eqnarray}    
\label{dipole}    
\hat{V}_{\cal E}^{ll^{\prime}}=    
\left( \begin{array}{ccc}    
                  k^{2} & \sqrt{2}k & 1  \\    
                  \sqrt{2}k & 1-k^{2}& -\sqrt{2}k \\    
                  1 &  -\sqrt{2}k & k ^{2}    
                  \end{array} \right)    
\frac{C^{ll^{\prime}}}{(1+k^{2})R^3},    
\end{eqnarray}    
whose coefficient $C^{ll^{\prime}}$, which is independent of both $R$  
and the electric field, is given by  
\begin{eqnarray}   
C^{ll^{\prime}} =     
-\mu^{2}   
([l][l'])^{1/2}   
\left( \begin{array}{ccc}      
                  l' & 2 & l \\      
                  0 & 0 &  0      
                  \end{array} \right)^{2}   
\frac{\Omega^{2} M_{F}^{2}(J(J+1)+F(F+1)-I(I+1))}{2(J(J+1)F(F+1))^{2}}     
\end{eqnarray}   
 Notice that the    
dipole-dipole interaction vanishes for s-waves, so that    
$\hat{V}_{\cal E}^{00}=0$.    
    
Within this simplified model we will refer to the scattering    
channels by the parities $\varepsilon_1$ and $\varepsilon_2$   
of the two molecules, along with the partial wave     
quantum number $l$.  Thus the incident channel for weak-field-seekers will    
be denoted $|\varepsilon_1 \varepsilon_2, l \rangle$ $=|--,0\rangle$.    
Recall that all other quantum numbers ($J$, $I$, $\Omega$, $F$, $M_F$)   
are assumed to have fixed values for each molecule.  
    
The explicit field-dependence in the coupling matrix (\ref{dipole})    
explains qualitatively the behavior of our ultracold weak-field-seeking    
molecules, which have incident quantum number $\varepsilon=-$.    
For zero electric field ($k=0$), there is no direct dipole-dipole coupling between  
identical molecules.  There is, however, an off-diagonal coupling to  
different channels with opposite parity,  
as can be seen in the form of the Hamiltonian (\ref{model}).    
This interaction    
brings in the dipole-dipole coupling in second order, contributing    
an effective dispersion-like potential $C_6^{eff}/R^6$, with a coefficient    
\begin{equation}    
\label{C6}    
C_6^{eff} = \frac{(C^{20})^{2}}{2\Delta}    
\propto \frac{{\mu}^{4}}{\Delta}   
\end{equation}  
for both s- and d- partial waves.   
For the OH molecule this effective coefficient is $\approx 4 \times 10^4$    
atomic units, far larger than for the alkali    
atoms that are familiarly trapped.  Thus, even in zero external field    
the effective interaction strength of polar molecules is quite large.    
This may imply the breakdown of the contact-potential approximation to describing    
Bose-Einstein condensates of polar molecules, even when their dipoles  
are not aligned by an external field \cite{Santos,You,Goral}.  We note that the     
quadrupole-quadrupole interaction is relatively unimportant, becoming  
larger than this effective dispersion interaction only when   
$R > \approx 3 \times 10^{5}$ a.u.   
    
When the field is switched on, the s-wave channels undergo a qualitative   
change.  Now the incident channel $|--,0 \rangle$ sees a direct coupling   
to its d-wave counterpart $|--,2\rangle$, via the matrix element    
$V^{l=0,l'=2} \propto \frac{k^{2}}{1+k^{2}} \frac{\mu^{2}}{R^{3}}$.    
This perturbation  generates an far stronger effective long-range potential   
of the form $C_4^{eff}/R^4$, with   
\begin{eqnarray}    
\label{slong}    
C_4^{eff}=  -(\frac{k^{2}}{1+k^{2}})^{2} \frac{\mu^{4}2m}{l(l+1)},   
\end{eqnarray}    
where $l=2$.   Thus the electric field is able to completely alter  
the character of the intermolecular interaction.  
    
For d-wave collisions, the dipole-dipole coupling is direct, but not in    
the limit of zero field, where the molecules are in parity eigenstates.    
At low fields ($k \ll 1$, where the Stark effect is quadratic),    
the diagonal coupling $V_{\cal E}^{22} \propto k^2/(1+k^2)$, is small.  
In this limit the molecules are nearly in parity eigenstates,    
hence do not ``know'' that they have dipole moments.  At larger fields  
this interaction grows in scale, thus ``activating'' the dipoles.     
This is why the rate constants shown in Figs 2,3 begin to evolve     
at fields near ${\cal E}={\cal E}_0$.  It is also why the contributions    
from all angular momentum projections except the one with ${\cal M}=0$   
contribute only weakly to scattering at low field.  The    
channels with ${\cal M} \ne 4$ are all of d-wave character, hence  
obey a threshold law $\sigma \propto  E^{2}$ at low fields,  then  
evolve to a $\sigma \propto {\rm const}$ threshold law at larger fields.   
    
\subsection{Large-field oscillations and long-range states of the    
[OH]$_2$ dimer}    
    
At fields larger than the critical field ${\cal E}_0$, the rate    
constants in Fig.2 exhibit oscillations with field.  Significantly, these occur    
only when the projection of total angular momentum ${\cal M}=4$,    
which is the only case in which s- and d- partial waves are mixed (Fig. 3).    
To understand this oscillating behavior of cross sections we     
show in Fig. 4(a) the adiabatic potential curves in the simplified 6-channel     
model~(\ref{model}).  In the case shown the field is ${\cal E}=10^4$ V/cm.    
In this figure a strong avoided crossing can be seen at $R \approx 60$ a.u.,    
corresponding to the crossing of the attractive $|\varepsilon_1  
\varepsilon_2,l \rangle=$  
$|--,0\rangle$ channel    
with the repulsive $|-+,2\rangle$ channel.  The strong dipole-dipole    
interaction between these different partial waves creates the     
adiabatic potential shown as a heavy black line and labeled $U_0$.  
    
This potential curve supports bound states of the [OH]$_2$ dimer.    
These bound states are of purely long-range character, similar to     
the long-range states of the alkali dimers \cite{Movre} that have been used  
in photoassociation spectroscopy studies of ultracold collisions    
\cite{Jones,Stwalley}.  Moreover, in the case of the [OH]$_2$    
states the shape of the potential $U_0$, hence the energies of the bound states,    
are strongly subject to the strength of the applied electric field.    
The curve in Fig. 4(a) in fact possesses no bound states in zero    
field, but five by the time the field reaches $10^4$ V/cm.      
More realistic adiabatic potentials are of course more elaborate,    
as shown in Fig. 4(b) for the more     
complete Hamiltonian that includes hyperfine levels.  Nevertheless,    
in this figure, too, can be seen adiabatic potential wells    
that will support bound states.    
    
The significance of these curves is twofold: the crossing is    
very adiabatic, implying that coupling to lower-energy channels    
is weak, and that therefore the cross sections depend only weakly    
on details of the short-range potentials.  This we have indeed     
verified by altering the short-range potential in the full calculation.    
    
Additionally, as the field strength grows and the potential becomes    
deeper, new bound states are added to the potential, causing    
scattering resonances to appear.  This is the cause of    
the oscillations observed in the rate constants in Fig. 2.  To    
illustrate this, we reproduce in Fig. 5 the complete elastic    
scattering rate constant (solid line), along with the    
same quantity as computed in the simple six-channel model    
(dashed line).  The qualitative behavior is nearly the same, namely,  
oscillations appear at fields above ${\cal E}_0$.  Moreover,  
the arrows in the figure indicate the values of the field for which    
a bound state of $U_0$ coincides with the scattering threshold.  
These fields correspond fairly well    
to the peaks, although they are somewhat offset by coupling to    
other channels.  Nevertheless, this simple picture  
clearly identifies the origin of these oscillations with the  
existence of long-range bound states.   
    
These resonant states are not Feshbach resonances, since there is    
no excitation of internal states of the molecules; nor are they    
shape resonances in the usual sense, since there is no barrier    
through which the wave function must tunnel.  Instead, they are the  
direct result of altering the interaction potential to place a  
bound state exactly at threshold \cite{landau}.    
Probing these states through direct scattering of weak-field-seeking    
states should reveal details about the long-range OH-OH interaction,    
making possible a comparison of theory and experiment without    
the need to fully understand the short-range [OH]$_2$ potential energy  
surface.

\section{Conclusion}      
In this paper we theoretically investigated ultracold collisions of  
ground state polar diatomic molecules in an electrostatic field,       
taking OH molecules as a prototype.  Focusing on weak-field-seeking  
states, we have strengthened the suppositions in Ref. \cite{bohnpolar}  
that long-range dipolar interactions drive inelastic scattering processes  
that are generally unfavorable for evaporative cooling of this species.  
However, at electric fields above a characteristic value ${\cal E}_0$,  
oscillations occur in both elastic and inelastic collision rates,  
implying that a regime may be found where the ratio  
$K_{elastic}/K_{inelastic}$ is favorable for cooling.  
Even though evaporative cooling may be difficult, the inelastic rates  
may nevertheless prove useful diagnostic tools for cold collisions  
of these molecules.  The Stark slowing technique  provides  
a means of launching a bunch of molecules toward a stationary trapped  
target, i.e., of making a real scattering experiment \cite{Meijer3,Ye}.  
  
For actual trapping and cooling purposes, for instance as a means of  
producing molecular Bose-Einstein condensates or degenerate Fermi gases, it  
seems likely that the molecules must be trapped in their strong-field  
seeking states.  Collisions of these species will present their own  
difficulties, since they will depend strongly on the short-range  
part of the potential energy surface.  However, the scattering length  
for OH-OH scattering may be determined by photoassociation spectroscopy  
to the long-range bound states we have described above.  This will be  
analogous to the determination of alkali scattering lengths   
\cite{Jones,Stwalley}, except that microwave, rather than optical, photons  
will be used.  The detailed properties of the long-range [OH]$_2$ states,   
and prospects for using  them in this way, therefore deserve further attention.  
      
This work was supported by the National Science Foundation. We acknowledge
useful discussions with J. Hutson and G. Shlyapnikov.

\newpage     
\begin{figure}     
\label{stark}     
\centerline{\includegraphics[width=0.50\linewidth,height=0.60\linewidth,angle=-90]{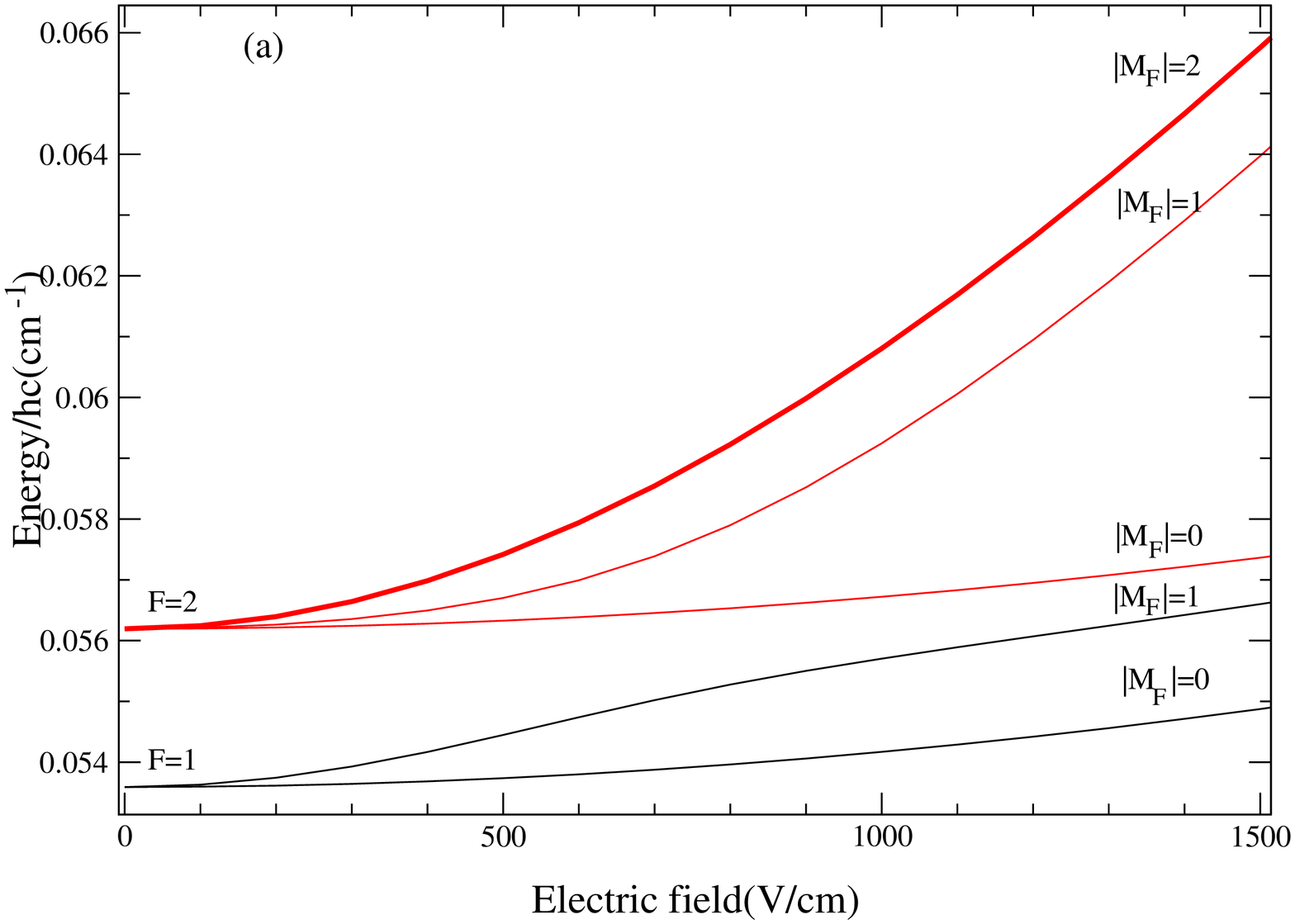}}     
\centerline{\includegraphics[width=.50\linewidth,height=0.60\linewidth,angle=-90]{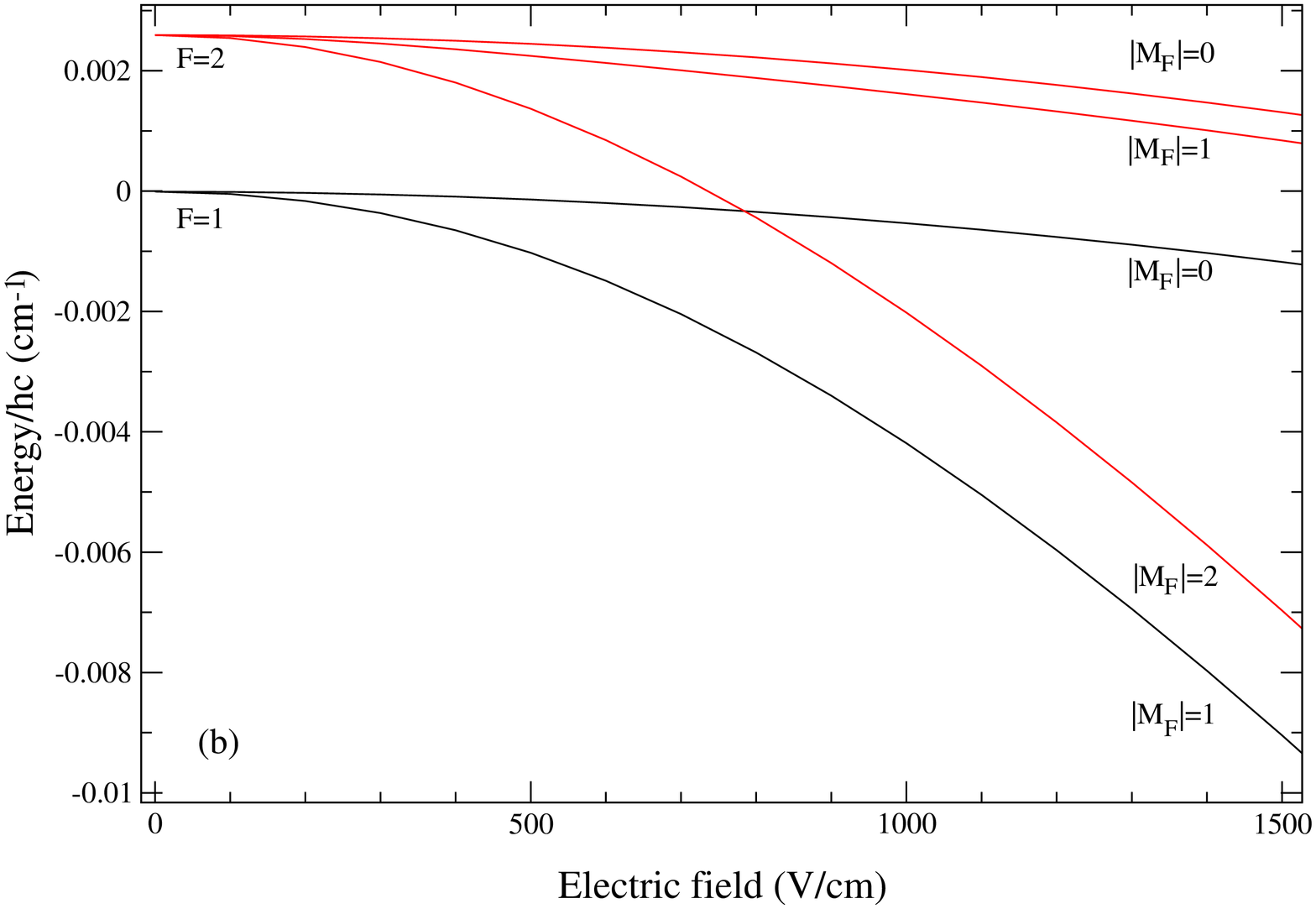}}     
\caption{     
The Stark effect in ground state OH molecules, taking into account hyperfine splitting.  
(a) shows the states that have odd parity $\varepsilon=-$ in zero electric field   
(f states), whereas (b) shows those of even parity (e states).  
The weak-field-seeking state with quantum numbers $F=M_F=2$,  the subject 
of this paper, is indicated by the heavy solid line.    Note that states with
$M_F = \pm |M_F|$ are degenerate in an electric field.
}     
\end{figure}     
     
\begin{figure}     
\label{rates}     
\centerline{\includegraphics[width=0.50\linewidth,height=0.60\linewidth,angle=-90]{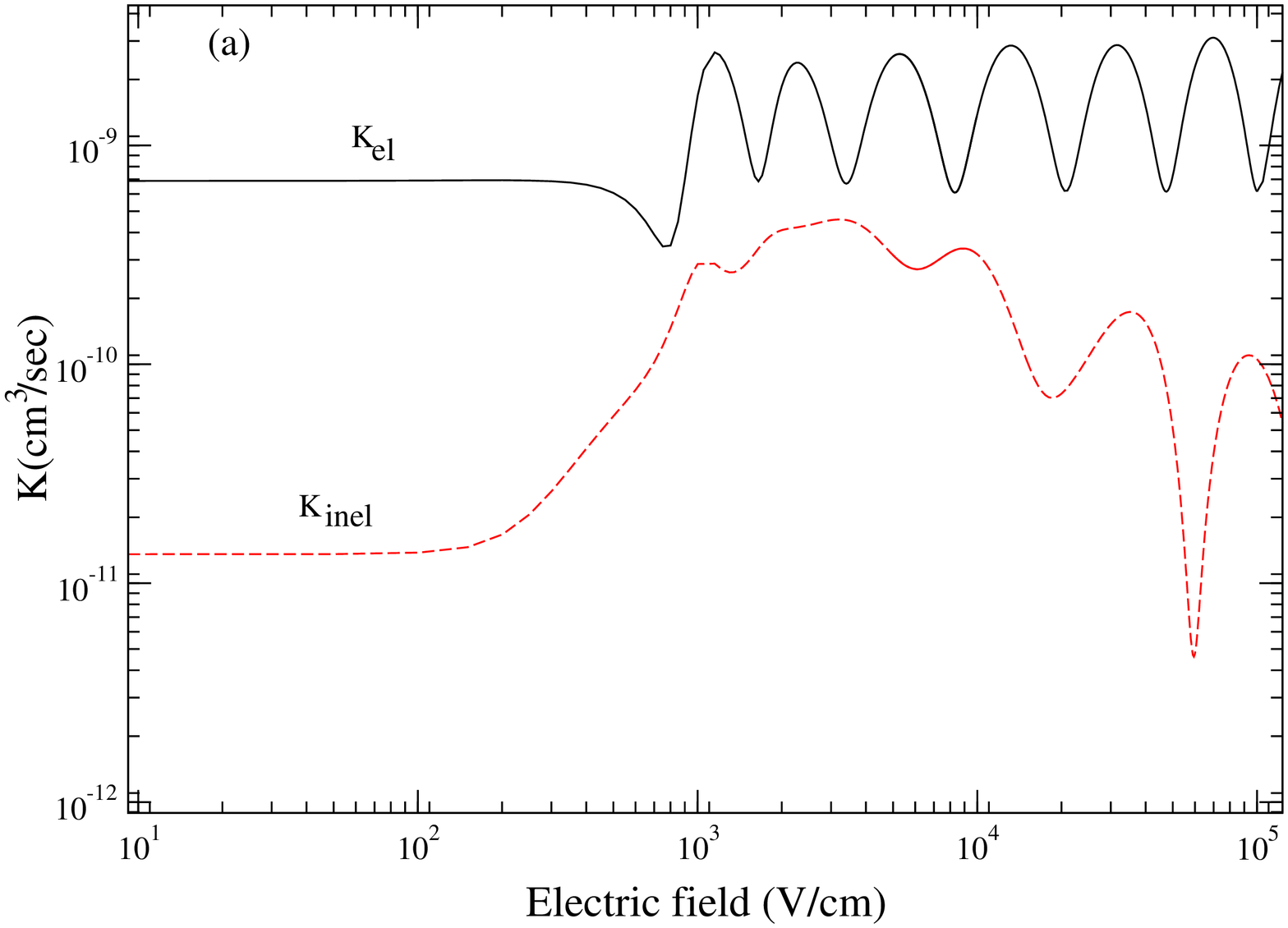}}     
\centerline{\includegraphics[width=.50\linewidth,height=0.60\linewidth,angle=-90]{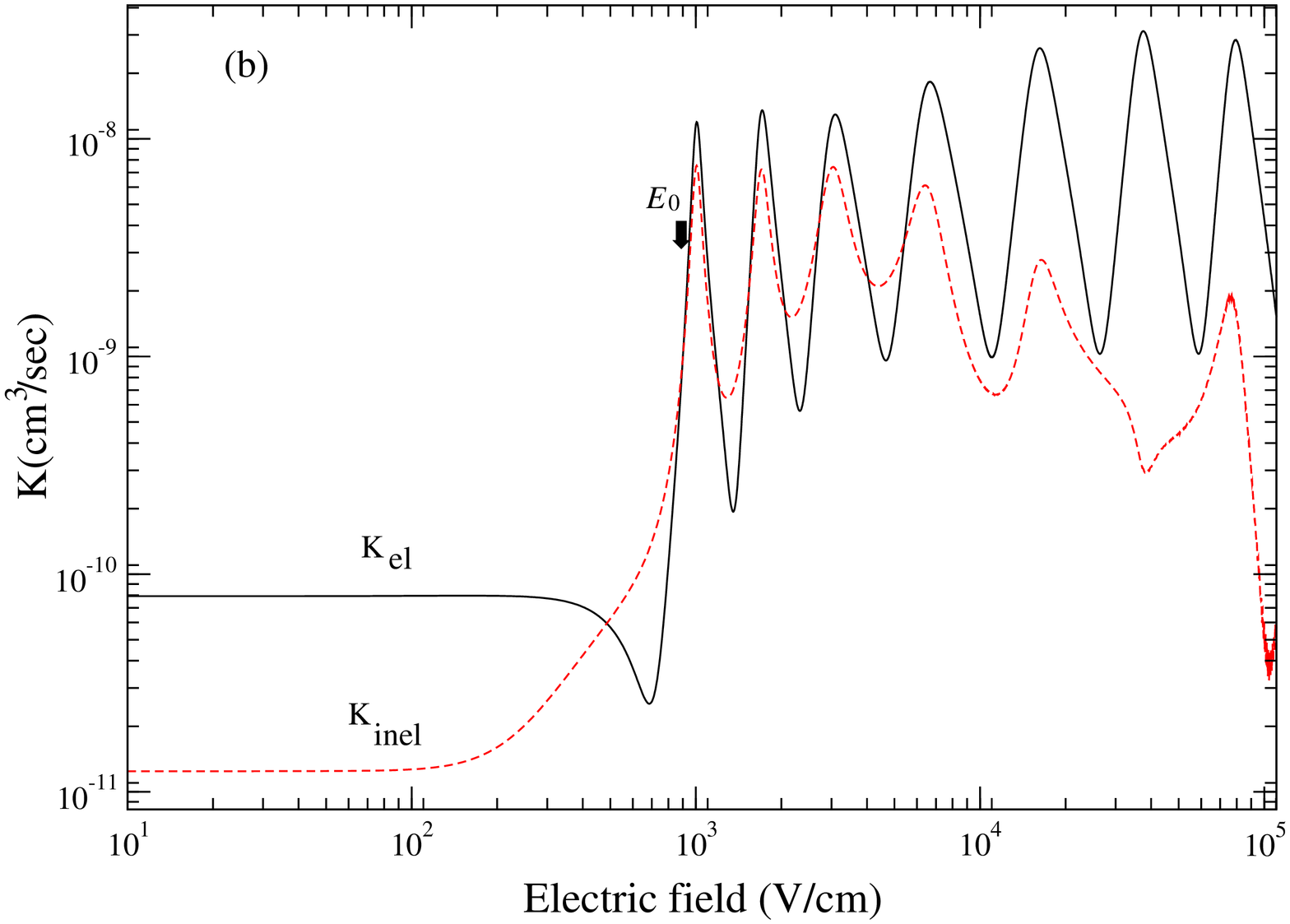}}     
\caption{     
Rate constants versus electric field for OH-OH collisions with molecules   
initially in their  $|FM_F,\varepsilon \rangle$ $=|22->$ state.  
Shown are the collision energies $100\mu K$ (a) and $1\mu K$(b).    
Solid lines denote elastic scattering rates, while dashed lines 
denote rates for inelastic collisions, in which one or both molecules
changes its internal state.
These rate constants exhibit characteristic oscillations in field when  
the field exceeds a critical field ${\cal E}_0 \approx 1000$ V/cm.  
}     
\end{figure}     
     
\begin{figure}     
\label{elastic}     
\centerline{\includegraphics[width=0.50\linewidth,height=0.60\linewidth,angle=-90]{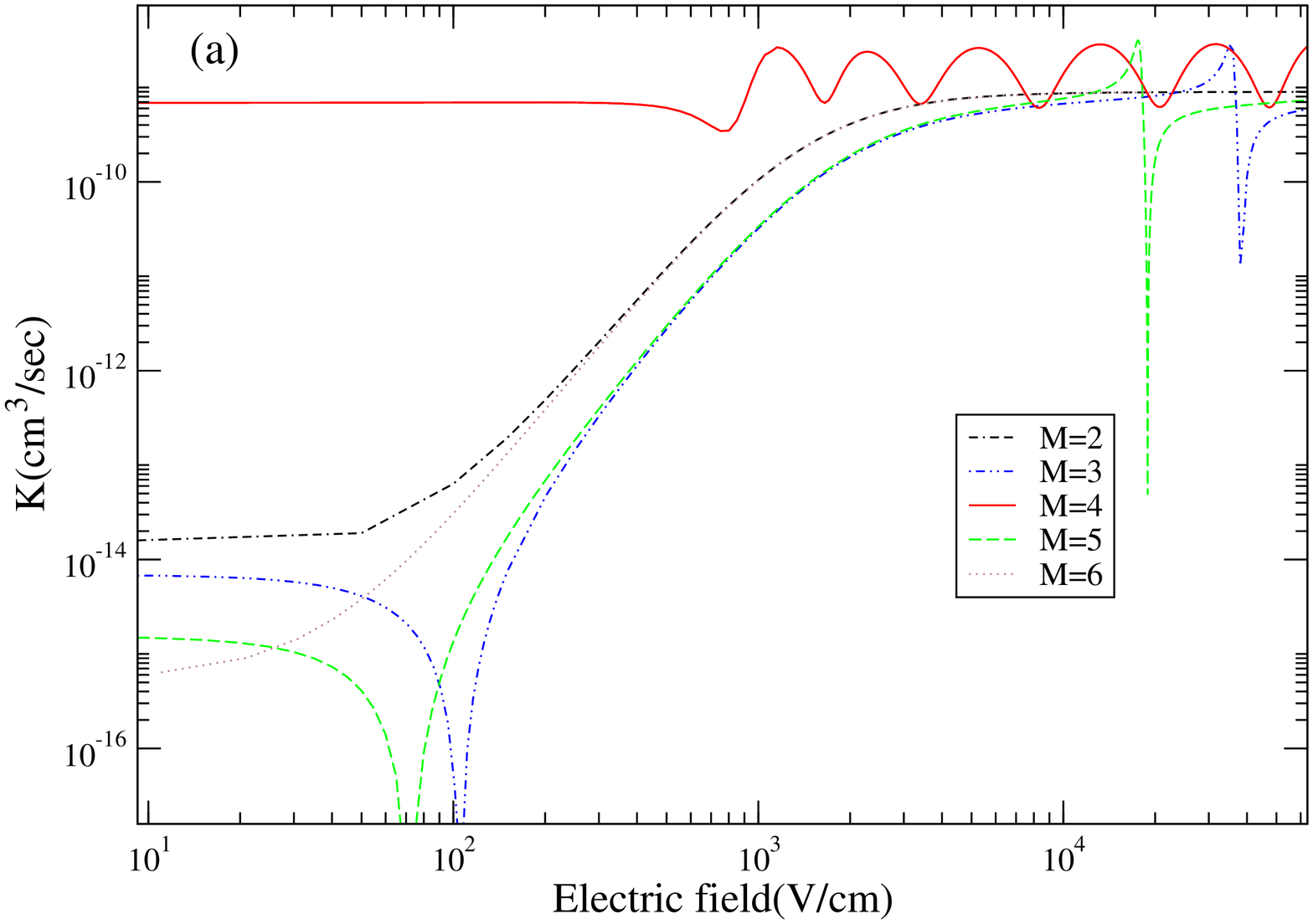}}     
\centerline{\includegraphics[width=0.50\linewidth,height=0.60\linewidth,angle=-90]{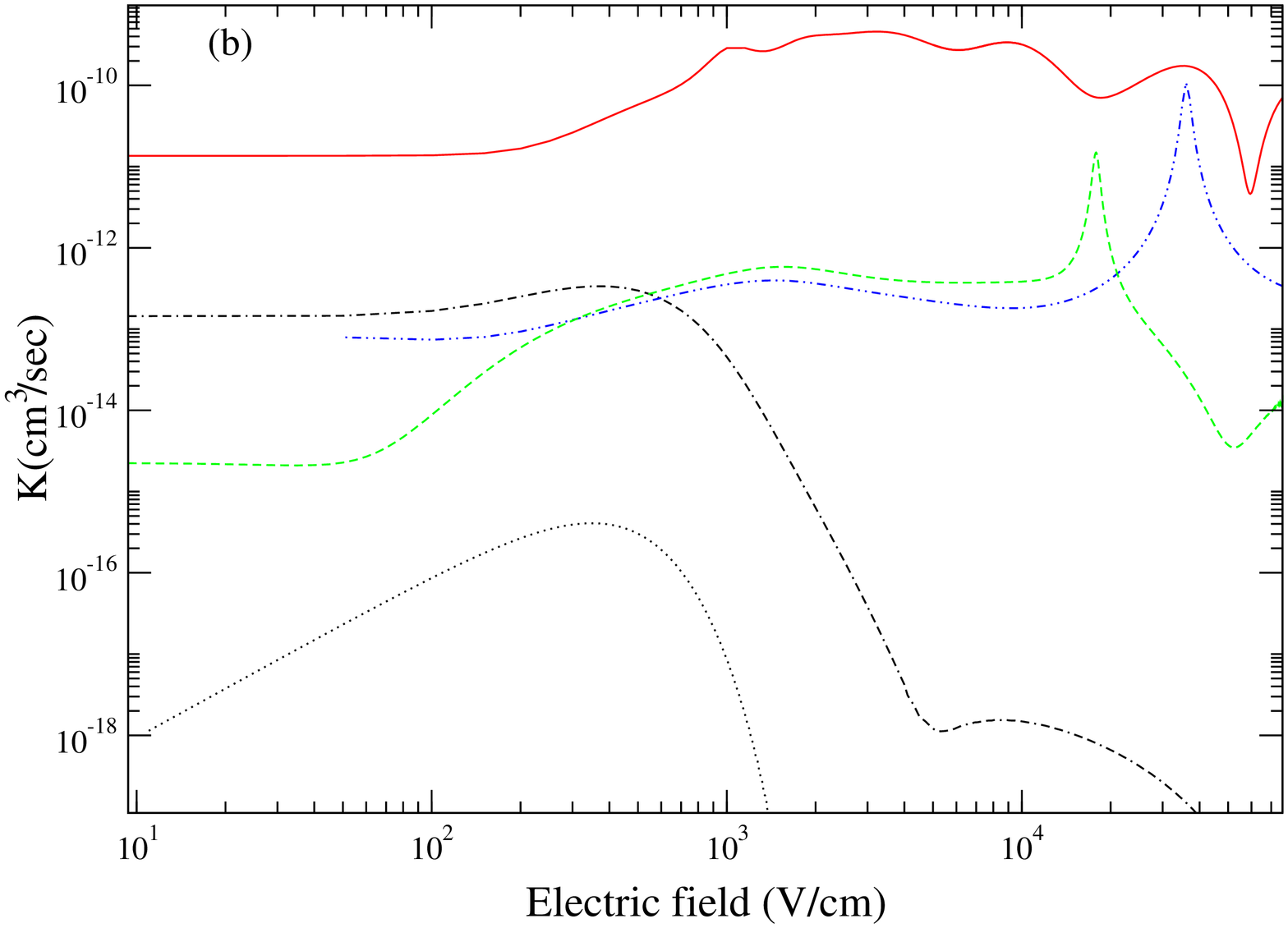}}     
\caption{     
Elastic (a) and inelastic (b) rate constants versus electric field for the      
same circumstances as in Figure 2(a).  The rates are separated into
contributions from different values of ${\cal M}$, the projection of
total angular momentum on the laboratory $z$-axis.
}     
\end{figure}

\begin{figure}     
\label{adiabatic}     
\centerline{\includegraphics[width=0.50\linewidth,height=0.60\linewidth,angle=-90]{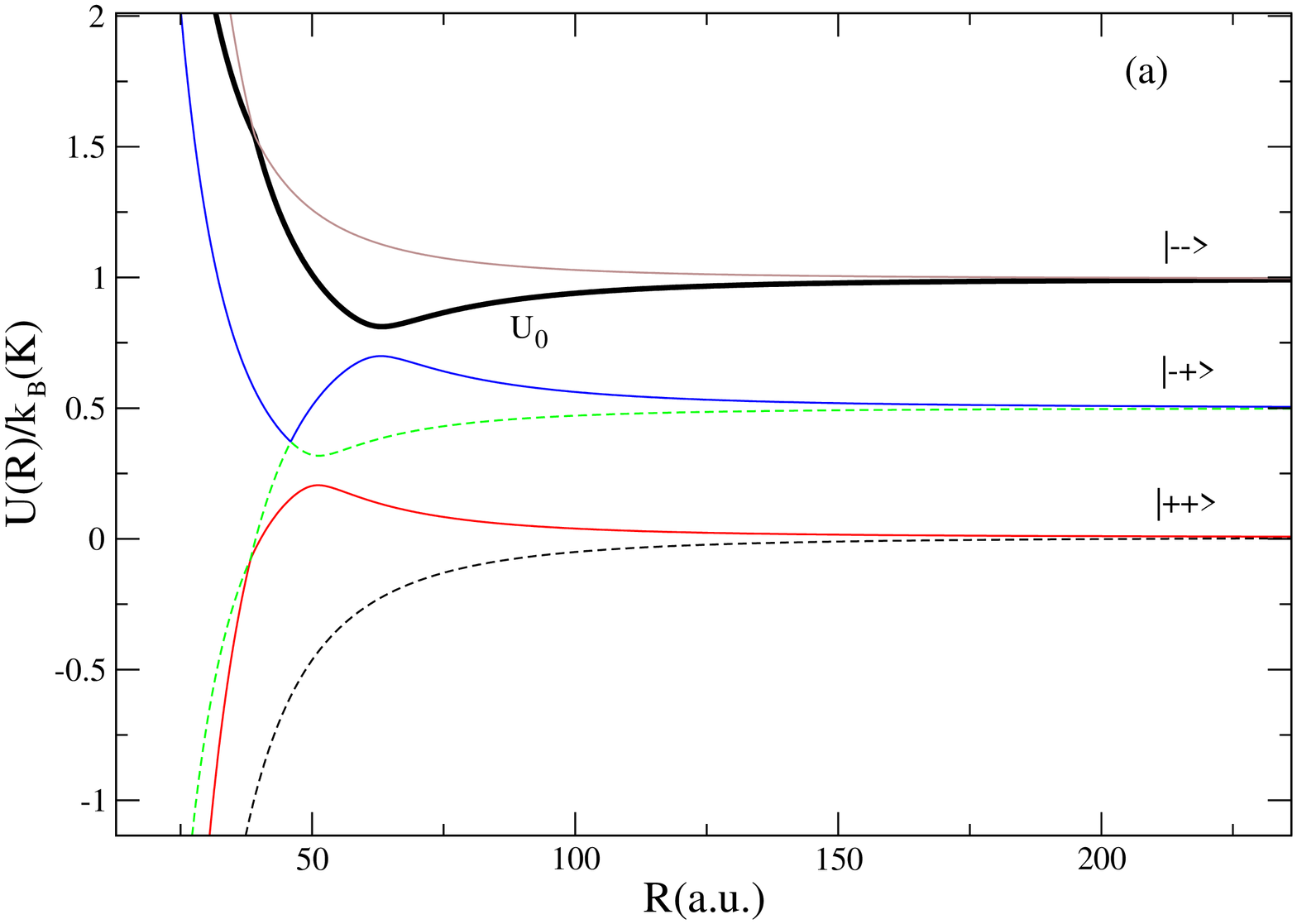}}     
\centerline{\includegraphics[width=0.50\linewidth,height=0.60\linewidth,angle=-90]{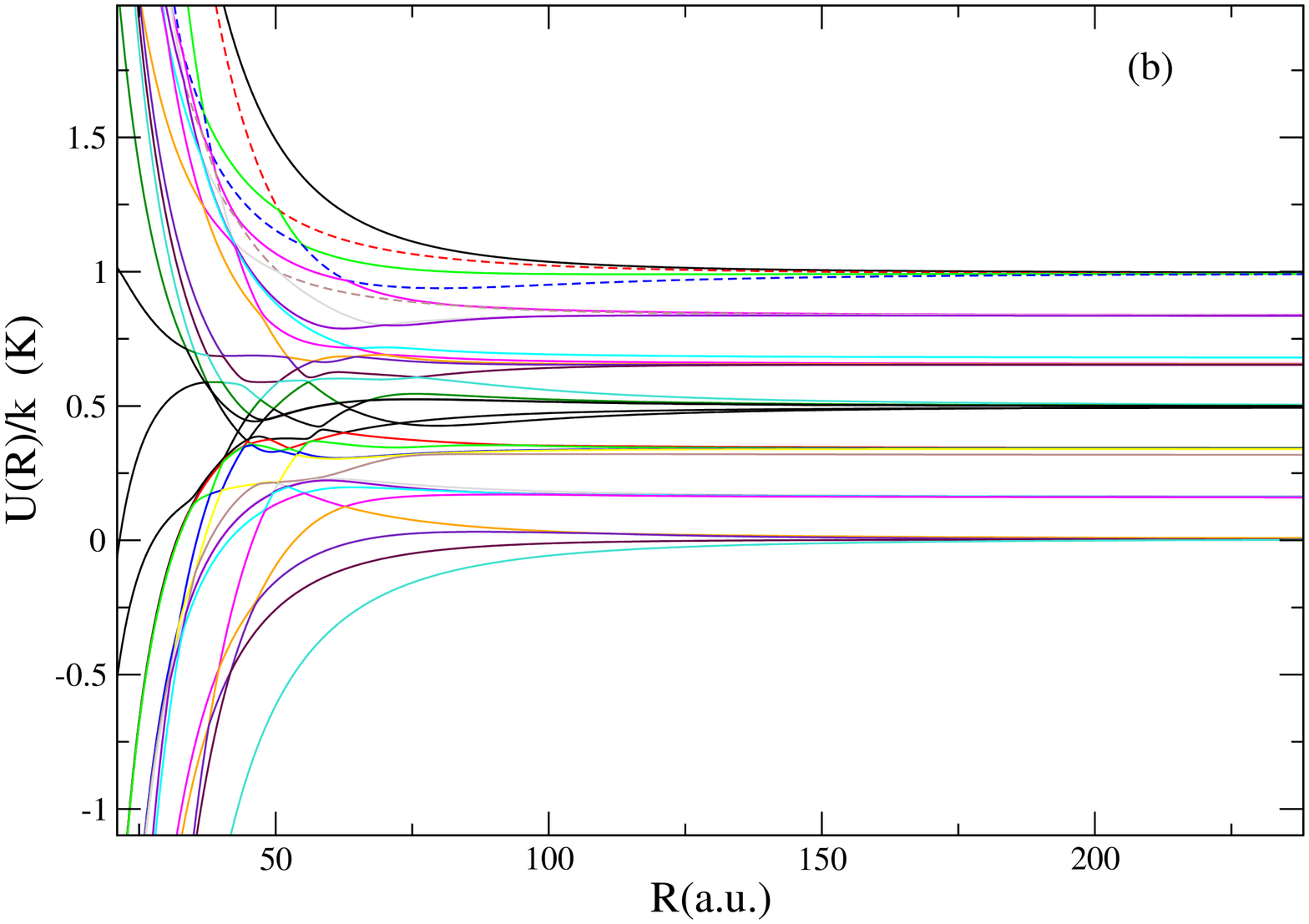}}     
\caption{     
Adiabatic potential energy curves.  The curves in (a) correspond to the  
simplified six-channel model described in the text, and show a long-range  
potential well (labeled $U_0$) that can hold bound states of the [OH]$_2$  
dimer.  The curves in (b) are those for the more complete calculation that  
includes hyperfine structure.  
}     
\end{figure}

\begin{figure}  
\label{bound}  
\centerline{\includegraphics[width=0.50\linewidth,height=0.60\linewidth,angle=-90]{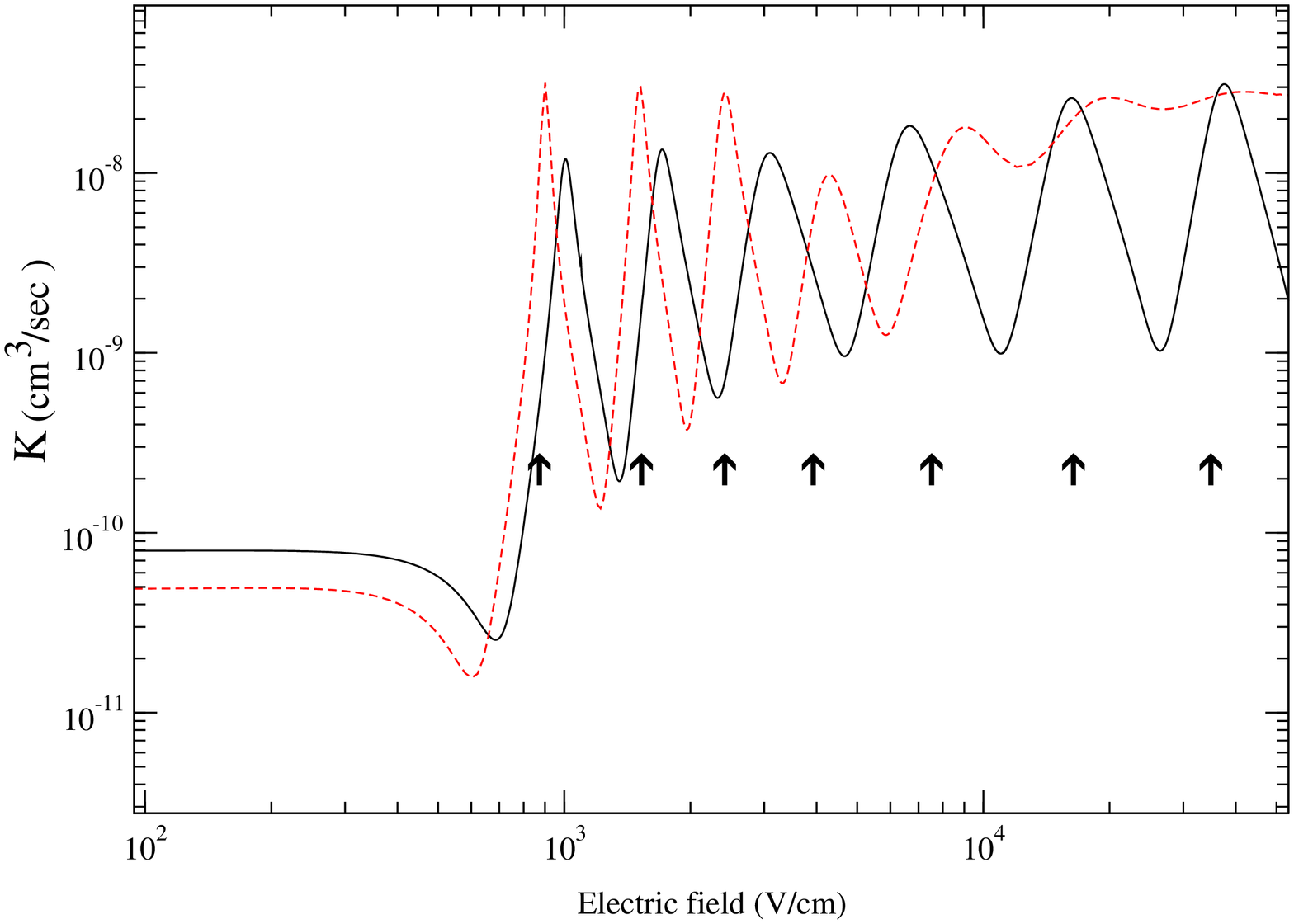}}  
\caption{  
Elastic rate constants versus field, as in Figure 2.  
The solid line reproduces the elastic rate constant from figure 2.  The  
dashed line is an approximate elastic rate constant based on the simplified  
six-channel model described in the text.  The arrows indicate values of  
the electric field at which bound states of the long-range potential  
$U_0$ (Figure 4) coincide with the scattering threshold.  
}  
\end{figure}  
     

\begin{references}      
  
 \bibitem{Santos} L. Santos, G.V. Shlyapnikov, P. Zoller and M. Levenstein,     
 Phys. Rev. Lett. {\bf 85}, 1791 (2000).;\\ Stefano Giovanazzi et.al.,  J.Phys.B: At.Mol.Opt.Phys. {\bf 34}, 4757 (2001)     
      
\bibitem{You} S. Yi and L. You, Phys. Rev. A {\bf 63}, 053607 (2001).  
  
\bibitem{Goral} K. G\'{o}ral, K. Rz\c{a}\.{z}ewski, and T. Pfau, Phys.  
Rev. A {\bf 61}, 051601 (2000); J.-P. Martikainen, M. Mackie, and  
K.-A. Suominen, Phys. Rev. A {\bf 64}, 037601 (2001).  
  
\bibitem{Shlyapnikov} M. A. Baranov, M. S. Mar'enko, V. S. Rychkov, and  
G. V. Shlyapnikov, cond-mat 0109437 (2001).  
  
\bibitem{DeMille} D. DeMille, quant-ph/0109083 (2001).  
  
\bibitem{Shaffer} J. P. Shaffer, W. Chalupczak, and N. P. Bigelow,  
Phys. Rev. Lett. {\bf 82}, 1124 (1999).  
  
\bibitem{Schloder} U. Schl\"{o}der, C. Silber, and Z. Zimmerman, Appl. Phys.  
B {\bf 73}, 801 (2001).  
  
\bibitem{Doyle} J. D. Weinstein, R. deCarvalho, T. Guillet, B.  
Friedrich, and J. M. Doyle, Nature {\bf 395}, 148 (1998).   
  
\bibitem{Egorov} D. Egorov,  
J. D. Weinstein, D. Patterson, B. Friedrich, and J. M. Doyle,  
Phys. Rev. A {\bf 63}, 030501 (2001).  
      
\bibitem{Meijer1} H. L. Bethlem, G. Berden, A. J. van Roij, F. M. H.  
Crompvoets, and G. Meijer, Phys. Rev. Lett. {\bf 84}, 5744 (2000).      
      
\bibitem{Meijer2} H. L. Bethlem, G. Berden, F. M. H. Crompvoets,      
R. T. Jongma, A. J. A. van Roij, and G. Meijer, Nature {\bf 406},      
491 (2000).      
  
\bibitem{Meijer3} H. L. Bethlem, F. M. H. Crompvoets, R. T. Jongma,  
S. Y. T. van der Meerakker, and G. Meijer, to appear in Phys. Rev. A.  
     
\bibitem{bohnpolar} John L. Bohn, Phys.Rev.A {\bf 63}, 052714(2001)  
 
\bibitem{landau} L.D. Landau and E.M. Lifshitz, Quantum Mechanics: Non- Relativistic Theory   
(Nauka, Moscow, 1989, 4th ed.)    
  
\bibitem{Shakeshaft} R. Shakeshaft, J. Phys. B {\bf 5}, L115 (1972).  
  
\bibitem{Deb} B. Deb and L. You, Phys. Rev. A {\bf 64}, 022717 (2001).  
  
\bibitem{Ye} J. Ye, private communication.  
     
\bibitem{coxon} J.A. Coxon,  Can.J.Phys {\bf 58}, 933 (1980).;\\     
Thomas D. Varberg and Kenneth M. Evenson, J.Mol.Spec. {\bf 157}, 55 (1993).     
     
\bibitem{kuhn} Bernd Kuhn  et.al., J.Chem.Phys. {\bf 111} 2565 (1998).  
      
\bibitem{harding} Lawrence B. Harding, J.Chem.Phys. {\bf 95} 8653 (1991).      
  
\bibitem{Bala} N. Balakrishnan and A. Dalgarno, Chem. Phys. Lett.    
{\bf 341}, 652 (2001).    
  
\bibitem{Johnson} B. R. Johnson, J. Comput. Phys. {\bf 13}, 445 (1973).  
    
\bibitem{mizushima}  M.Mizushima, The Theory of Rotating Diatomic Molecules (Wiley, New York, 1975);\\     
Helene Lefebvre- Brion, Robert W. Field, Perturbations in the Spectra of Diatomic Molecules      
(Academic Press Inc.(London) Ltd., 1986);\\     
Karl F. Freed, J.Chem.Phys. {\bf 45} 4214 (1966).     
      
\bibitem{miller} Steven M. Miller and David C. Clary, J.Chem.Phys. {\bf 98} 1843 (1993).      
      
\bibitem{schreel} K.Schreel and J.J.Meulen, J.Phys.Chem.A {\bf 101}, 7639 (1997).      
     
\bibitem{avoird} Ad van der Avoird, Topics in Curr. Chem, {\bf 93} 1 (1980)      
  
\bibitem{Roberts} J. L. Roberts, J. P. Burke, S. L. Cornish, N. R. Claussen,   
E. A. Donley, and C. E. Wieman, Phys. Rev. A {\bf 64}, 024702 (2001).  
  
\bibitem{Loftus} T. Loftus, C. A. Regal, C. Ticknor, J. L. Bohn, and D. S. Jin,  
submitted to Phys. Rev. Lett.  
  
\bibitem{Avdeenkov} Alexandr V. Avdeenkov and John L. Bohn, Phys.Rev.A {\bf 64}, 052703(2001)      
      
\bibitem{Volpi} A. Volpi and J. L. Bohn, to appear in Phys. Rev. A.  
  
\bibitem{Movre} M. Movre and G. Pichler, J. Phys. B {\bf 10}, 2631 (1977);  
W.C. Stwalley, Y.-H. Uang, and G. Pichler, Phys. Rev. Lett.  
{\bf 41}, 1164 (1978).  
  
\bibitem{Jones} K. M. Jones, P. S. Julienne, P. D. Lett, W. D. Phillips,   
E. Tiesinga, and C. J. Williams, Europhys. Lett. {\bf 35}, 85 (1996).  
  
\bibitem{Stwalley} J. P. Burke, Jr. {\it et al}, Phys. Rev. A {\bf 60},  
4417 (1999); C. J. Williams {\it et al.}, Phys. Rev. A {\bf 60}, 4427 (1999).  
       
\end{references}
\end{document}